\documentclass[12pt]{iopart}

\usepackage{graphicx}
\usepackage{csquotes}
\MakeOuterQuote{"}
\usepackage{bm}
\usepackage{hyperref}
\usepackage[numbers,square,sort&compress]{natbib}
\usepackage{color}

\begin{document}

\title{The effects of kinetic instabilities on the electron cyclotron emission from runaway electrons}

\author{Chang Liu}
\address{Princeton Plasma Physics Laboratory, Princeton, New Jersey 08540, USA}
\ead{cliu@pppl.gov}
\author{Lei Shi}
\address{University of California, Irvine, California 92697, USA}	
\author{Eero Hirvijoki}
\address{Princeton Plasma Physics Laboratory, Princeton, New Jersey 08540, USA}
\author{Dylan P. Brennan}
\address{Princeton University, Princeton, New Jersey 08544, USA}
\author{Amitava Bhattacharjee}
\address{Princeton Plasma Physics Laboratory, Princeton, New Jersey 08540, USA}
\address{Princeton University, Princeton, New Jersey 08544, USA}
\author{Carlos Paz-Soldan}
\address{General Atomics, San Diego, California 92186, USA}
\author{Max E. Austin}
\address{University of Texas, Austin, Texas 78712, USA}

\begin{abstract}
In this paper we show that the kinetic instabilities associated with runaway electron beams play an essential role for the production of high-level non-thermal electron-cyclotron-emission (ECE) radiation. Most of the non-thermal ECE comes from runaway electrons in the low-energy regime with large pitch angle, which are strongly scattered by the excited whistler waves. The power of ECE from runaway electrons is obtained using a synthetic diagnostic model based on the reciprocity method. The electron distribution function is calculated using a kinetic simulation model including the whistler wave instabilities and the quasilinear diffusion effects. Simulations based on DIII-D low-density discharge reproduces the rapid growth of the ECE signals observed in DIII-D experiments. Unlike the thermal ECE where radiation for a certain frequency is strongly localized inside the resonance region, the non-thermal ECE radiation from runaway electrons is nonlocal, and the emission-absorption ratio is higher than that of thermal electrons. The runaway electron tail is more significant for ECE with higher frequencies, and the ECE spectrum becomes flatter as RE population grows. The nonlinear behavior of the kinetic instabilities is illustrated in the osculations of the ECE waves. The good agreement with the DIII-D experimental observations after including the kinetic instabilities clearly illustrate the significance of the scattering effects from wave-particle interactions, which can also be important for runaway electrons produced in disruptions.  
\end{abstract}
\submitto{\NF}

\maketitle

\section{Introduction}

In recent tokamak studies, runaway electrons (RE) have drawn a lot of attention due to their importance in disruptions\cite{boozer_pivotal_2018}. The classic theory of runaway electrons is based on the dominance of a strong electric field force over the collisional force in the high-energy regime, where an electron an electron can be accelerated to extremely high energy if a threshold in energy is exceeded\cite{dreicer_electron_1959,dreicer_electron_1960,connor_relativistic_1975}. Based on this theory, it is predicted that in a typical disruption event in ITER, the strong inductive electric field can cause an avalanche growth\cite{rosenbluth_theory_1997} of a high-energy RE population up to tens of MeV, which can bring severe damage to the plasma facing material\cite{lehnen_disruption_2011}. Given the importance of RE physics, many experiments have been conducted in current tokamaks\cite{granetz_itpa_2014}, including flattop scenarios\cite{paz-soldan_growth_2014} and disruption cases\cite{fredrickson_control_2015} , to validate the theoretical model and provide predictions for ITER. In these experiments, the dynamics of runaway electrons are diagnosed mainly through the radiation emitted, including the hard X-ray (HXR) and gamma rays from bremsstrahlung\cite{paz-soldan_non-thermal_2016}, visible and infrared lights from synchrotron emission (SE)\cite{hoppe_interpretation_2017}, and radio-frequency waves from  electron cyclotron emission (ECE). Analysis of the diagnostic results show that, in addition to the electric field force and the collision force, the radiation reaction force\cite{stahl_effective_2015} and the kinetic instabilities like whistler waves\cite{aleynikov_stability_2015} are also very important in the RE dynamics, making the runaway electron distribution very different from what classic theory predicts.

The ECE has been widely used in tokamak experiments to diagnose the electron temperature ($T_{e}$)\cite{bornatici_electron_1983}. The diagnostic is based on the radiation of electron cyclotron waves (ECWs), as a consequence of the gyro-motion of electrons around magnetic field lines. ECWs can be emitted and absorbed by the electrons when the resonance condition is satisfied. Depending on the absorption coefficients, ECWs with different frequencies and polarizations have different optical thicknesses with respect to the plasma. According to the Kirchhoff's law of thermal radiation, if the plasma is locally in thermal equilibrium and optically thick so that the wave absorption in the resonance region is strong, then the radiated power from the resonance region is proportional to the local electron temperature. Based on this mechanism, the ECE-imaging (ECEI) approach, which scans the ECE signals with various frequencies and propagation angles, has been used to obtain the spatial profile and fluctuations of electron temperature.

The high-energy runaway electrons can also produce ECE during gyro-motions. In many RE experiments in tokamaks including both the flattop\cite{paz-soldan_growth_2014, zhou_investigation_2013} and disruptions scenarios\cite{fredrickson_control_2015}, ECE signals have been observed that are much more intense than the ECE from thermal electrons.  Compared to HXR or gamma rays which mainly depend on the electron energy, the ECE power depends on both the energy and the pitch angle of the runaway electrons, similar to SE. On the other hand, the ECE has a wide-angle distribution which is rather different from the strong forward-beaming of SE. 
Given that the RE tail is not in thermal equilibrium and the absorption of ECWs is much smaller than that of thermal electrons, the ECE radiation power is not primarily determined by the electron temperature in the manner specified by Kirchoff's law. The ECE from suprathermal electrons has been studied in lower hybrid current drive experiments\cite{mazzucato_absorption_1985,kato_nonthermal_1986,preische_radially_1996} using the Doppler broadening of the ECE spectrum, but for RE tail with a wide energy distribution, a synthetic diagnostic model is required to interpret the more complicated non-thermal ECE signals that elucidate the electron distribution function.

In this paper, we present an ECE synthetic diagnostic model for arbitrary electron distributions. The model is applied to study relativistic RE tail including the kinetic instabilities. Unlike the forward method used in previous codes like HORACE\cite{harvey_electron_1993}, in our model we use a backward method to calculate the ECE power based on the reciprocal theorem\cite{piliya_application_2002}, an innovative approach which separates the wave emission and absorption by introducing an artificial wave propagating against the radiation direction along the same ray path. The reflection loss and the polarization scrambling\cite{costley_electron_1974,hutchinson_electron_1977,sillen_electron_1986} of the waves are included in the reciprocal calculations. The RE distribution function is calculated using a newly-developed kinetic simulation model, coupled to a kinetic instability calculation framework using the quasilinear diffusion model\cite{liu_role_2018}. The results show that, when kinetic instabilities are taken into account, the synthetic diagnostic shows good agreement with experiments for both X-wave and O-wave, including the growth rate and peak amplitude of radiation power\cite{paz-soldan_non-thermal_2016}, the evolution of the ECE spectrum shape, and the oscillatory behavior of ECE signals observed in RE decaying stage\cite{spong_first_nodate}. The radiation power is further expressed using the ECE weight function, which shows that most of the non-thermal ECE radiation comes from runaway electrons with low-energy and large pitch angle, and the excitation of whistler waves by REs plays an essential role in producing this radiation.

This paper is organized as follows. In Sec. \ref{sec:ECW} we introduce the ECE synthetic diagnostic model, which is based on the reciprocity theorem in electrodynamics. In Sec. \ref{sec:RE-simulation} the kinetic simulation which provides the RE distribution for the synthetic diagnostic is introduced. Then in Sec. \ref{sec:ECE-simulations} the synthetic ECE signals obtained using these two models based on a DIII-D low-density flattop RE experiment is shown, which reproduces features observed in experiments. In Sec. \ref{sec:explanation}, we explain the physics mechanism causing the strong non-thermal ECE from RE in our simulation results, and calculate the ECE weight function in this case. In Sec. \ref{sec:conclusions} we summarize. 

\section{Absorption and emission of electron cyclotron wave}
\label{sec:ECW}

To calculate the wave power radiated by the plasma and received by a specified antenna, one can either use the forward or the backward method. In the forward method, the radiation intensity at every position in plasma is solved through the radiation transport equation\cite{harvey_electron_1993}, which is derived from the wave kinetic equation following the Wentzel-Krammers-Brillouin (WKB) method. In solving the transport equation, one needs to calculate an integral along the ray path, and both the emission and the absorption of the wave needs to be taken into account. On the other hand, in the backward method the absorption and emission of the wave are treated separately, which can facilitate the computation. In this method, one needs to calculate an artificial wave which propagates from the antenna back into the plasma, to get the dependence function of the radiation power on the emission of every point, based on the so-called reciprocal theorem\cite{piliya_application_2002}. It can be proved mathematically that the two methods are equivalent, and for a 1D radiation problem of the kind studied in the present paper, the calculations are not much different. However, for more complicated radiation problems such as ECEI, the backward method is more computationally advantageous since it does not need to include calculations of the radiation not entering the antenna.

We now focus on the application of the backward method to the ECE. According to the reciprocal theorem, to get the dependence function we need to calculate the electric fields of an artificial wave. This artificial wave has the same frequency and polarization but  propagates in a direction opposite to that of the ECWs. The fields of the artificial wave in plasma can be solved according to the Maxwell equation,
\begin{equation}
\label{eq:maxwell}
  \nabla\times \nabla \times \mathbf{E}+\frac{\omega^2}{c^{2}}\underline{\underline{\epsilon}}\cdot \mathbf{E}=\frac{4\pi}{c}\mathbf{j},
\end{equation}
where $\mathbf{E}$ is the wave electric fields, $\omega$ is the wave frequency, and $c$ is the speed of light. $\underline{\underline{\epsilon}}$ is the plasma dielectric tensor. $\underline{\underline{\epsilon}}$ can be separated into a Hermitian part $\underline{\underline{\epsilon}}^{H}$ and an anti-Hermitian part $\underline{\underline{\epsilon}}^{A}$. The Hermitian part determines the dispersion relation and the polarization of the wave, and the anti-Hermitian part determines the damping. In this work we use the cold electron dielectric tensor for the Hermitian part of $\underline{\underline{\epsilon}}$\cite{stix_waves_1992} ignoring the ion contribution and the thermal effect, assuming their correction to the dispersion relation is small. The anti-Hermitian part includes both the absorption of waves due to particle resonance ($\underline{\underline{\epsilon}}^{Ar}$) and damping from collisions($\underline{\underline{\epsilon}}^{Ac}$). The resonance absorption depends sensitively on the distribution function near the resonance point. The collision damping can be calculated from the electron collision frequency and the wave dispersion relation\cite{aleynikov_stability_2015}, which is independent of the distribution function. The current $\mathbf{j}$ on the right-hand-side of Eq. (\ref{eq:maxwell}) represents the emission of the wave from the fluctuation current, which can be ignored in the calculation of the artificial wave and will be addressed later.

For simplicity, we consider the artificial wave that is propagating along the major radius direction only, propagating from the low-field-side to the high-field. Set $x$ as the direction of the wave propagation, and $z$ as the magnetic field direction, the polarization of the X-wave is
\begin{equation}
\label{eq:Ex}
  E_{x}=\frac{i\omega_{ce}\omega_{pe}^{2}}{\omega(\omega^2-\omega_{ce}^{2}-\omega_{pe}^{2})}E_{y}-\frac{\omega^2-\omega_{ce}^{2}}{\omega^{2}-\omega_{ce}^{2}-\omega_{pe}^{2}}(\underline{\underline{\epsilon}}^{A}\cdot \mathbf{E})_{x},\quad E_{z}=0,
\end{equation}
where $\omega_{ce}$ and $\omega_{pe}$ are the electron cyclotron frequency (we choose $\omega_{ce}<0$) and plasma frequency obtained from the local density and magnetic field. Eq. (\ref{eq:maxwell}) can then be simplified as
\begin{equation}
  \label{eq:Ey-differential}
  \frac{\partial^{2} E_{y}}{\partial x^{2}}+\frac{(\omega^2-\omega_{pe}^{2})-\omega^2\omega_{ce}^{2}}{(\omega^2-\omega_{pe}^{2}-\omega_{ce}^{2})c^{2}}E_{y}+\frac{\omega^2}{c^{2}}\frac{i\omega_{ce}\omega_{pe}^{2}}{\omega(\omega^2-\omega_{ce}^{2}-\omega_{pe}^{2})}(\underline{\underline{\epsilon}}^{A}\cdot \mathbf{E})_{x}+\frac{\omega^2}{c^{2}}(\underline{\underline{\epsilon}}^{A}\cdot \mathbf{E})_{y}=0.
\end{equation}
Note in Eqs. (\ref{eq:Ex}) and (\ref{eq:Ey-differential}), the $\underline{\underline{\epsilon}}^{A}$  terms are much smaller than the rest of the terms. Taking the first order approximation, Eq. (\ref{eq:Ey-differential}) can be solved using the WKB approximation, and the result is
\begin{equation}
\label{eq:WKB}
  E_{y}=\frac{C}{\sqrt{k(x)}}\exp\left[\int dx \left[i k(x)+k_{i}(x)\right]\right],
\end{equation}
where $k(x)=\sqrt{\left[\left(\omega^2-\omega_{pe}^{2}\right)^{2}-\omega^2\omega_{ce}^{2}\right]/\left[\left(\omega^2-\omega_{pe}^{2}-\omega_{ce}^{2}\right)c^{2}\right]}$ is the real component of the wave vector. $k_{i}$ is the imaginary component,
\begin{equation}
\label{eq:ki}
  k_{i}(x)=\frac{\omega^2}{c^{2}}\frac{\left(\mathbf{E}^{\dagger}\cdot\underline{\underline{\epsilon}}^{A}\cdot \mathbf{E}\right)}{2 k(x) |E_y|^{2}},
\end{equation}
which describes the absorption of the X-wave due to particle resonances.

For O-wave, the polarization is
\begin{equation}
\label{eq:O-wave-polar}
E_{x}=0,\quad E_{y}=0,
\end{equation}
and $E_{z}$ satisfies the differential equation,
\begin{equation}
\label{eq:Ez-differential}
\frac{\partial^{2} E_{z}}{\partial x^{2}}+\frac{\omega^{2}-\omega_{pe}^{2}}{c^{2}}E_{z}+\frac{\omega^2}{c^{2}}(\underline{\underline{\epsilon}}^{A}\cdot \mathbf{E})_{z}=0.
\end{equation}
We can similarly derive the WKB equation for $E_{z}$,
\begin{equation}
\label{eq:WKB}
E_{z}=\frac{C}{\sqrt{k(x)}}\exp\left[\int dx \left[i k(x)+k_{i}(x)\right]\right],
\end{equation}
where
\begin{equation}
  k(x)=\sqrt{\frac{\omega^{2}-\omega_{pe}^{2}}{c^{2}}},
\end{equation}
\begin{equation}
\label{eq:ki}
k_{i}(x)=\frac{\omega^2}{c^{2}}\frac{\left(\mathbf{E}^{\dagger}\cdot\underline{\underline{\epsilon}}^{A}\cdot \mathbf{E}\right)}{2 k(x) |E_z|^{2}}.
\end{equation}
For the artificial wave, constant $C$ is arbitrary and here we choose $C=1$. The value of $\mathbf{E}$ along the ray can then be calculated by integrating Eq. (\ref{eq:WKB}).

When reaching the wall of tokamak, the artificial wave gets reflected into the opposite direction of the incoming ray. The intensity of the wave ($I_{X}=E_{y}^{2}k$ for X-wave and $I_{O}=E_{z}^{2}k$ for O-wave) will be multiplied by the reflection factor $\alpha_r<1$ due to the reflection power loss. In addition, the intensity of the two polarizations will transfer between each other due to the polarization scrambling, which can be characterized by a polarization transfer fraction $\alpha_p$. The intensity after a reflection $I'$ will thus follow the equations\cite{austin_electron_2003,austin_determination_1997},
\begin{equation}
  I'_{X}=\alpha_r \left[(1-\alpha_p)I_{X}+\alpha_p I_{O}\right]
\end{equation}
\begin{equation}
I'_{O}=\alpha_r \left[(1-\alpha_p)I_{O}+\alpha_p I_{X}\right]
\end{equation}

After obtaining the artificial wave electric fields $\mathbf{E}$, the total radiation power of the ECE with the same frequency can be calculated according to the reciprocity theorem, and sum over X and O waves,
\begin{equation}
\label{eq:reciprocal}
  P(\omega)=\sum_{X,O} \int dx \mathbf{E}^{\dagger}\cdot \underline{\underline{K}} \cdot \mathbf{E},
\end{equation}
where $\underline{\underline{K}}$ is the correlation tensor for the fluctuating current density, which represents the emission of ECWs by the resonant particles (the term on the right-hand-side of Eq. (\ref{eq:maxwell})). The reciprocal theorem proves that the power obtained from Eq. (\ref{eq:reciprocal}) is the same as the power calculated using the forward method, but in the backward method the absorption and emission are separately calculated and only coupled through Eq. (\ref{eq:reciprocal}).

The values of $\underline{\underline{\epsilon}}^{Ar}$ and $K$ can be calculated from the electron distribution function\cite{aleynikov_stability_2015,harvey_electron_1993,shi_synthetic_2016-1},
\begin{equation}
\label{eq:epsilonA}
  \underline{\underline{\epsilon}}^{Ar}=\frac{\omega_{pe}^{2}}{\omega}\int d^{3}\mathbf{p} \sum_{n=-\infty}^{\infty}\delta(\omega-k_{\parallel}v_{\parallel}-n\omega_{cb})\left[\frac{1}{v}\frac{\partial f}{\partial p}+\frac{\omega\cos\theta-k_{\parallel}v}{\omega p v_{\perp}}\right] \underline{\underline{T_{n}}},
\end{equation}
\begin{equation}
\label{eq:K}
  \underline{\underline{K}}=\frac{\omega_{pe}^{2}}{\pi}\int d^{3}\mathbf{p} \sum_{n=-\infty}^{\infty}\delta(\omega-k_{\parallel}v_{\parallel}-n\omega_{cb}) f \underline{\underline{T_{n}}},
\end{equation}
where
\begingroup
\renewcommand*{\arraystretch}{2.22}
\begin{equation}
\label{eq:Tn}
\arraycolsep=10pt
\underline{\underline{T_{n}}}=\left(\begin{array}{ccc}
\displaystyle\frac{n^{2}\omega_{cb}^{2}}{k_{\perp}^{2}}J_{n}^{2}&\displaystyle\frac{i n\omega_{cb}v_{\perp}}{k_{\perp}}J_{n}J'_{n}&\displaystyle\frac{n\omega_{cb}v_{\parallel}}{k_{\perp}}J_{n}^{2}\\
\displaystyle\frac{-i n\omega_{cb}v_{\perp}}{k_{\perp}}J_{n}J'_{n}&\displaystyle v_{\perp}^{2}J'^{2}_{n}&-i v_{\perp} v_{\parallel}J_{n}J'_{n}\\
\displaystyle\frac{n\omega_{cb}v_{\parallel}}{k_{\perp}}J_{n}^{2}&i v_{\perp} v_{\parallel}J_{n}J'_{n}&v_{\parallel}^{2} J_{n}^{2}
\end{array}\right),
\end{equation}$k_{\parallel}$ is the wave vector parallel to magnetic field (in our ECE calculations $k_{\parallel}=0$), and $k_{\perp}$ is the perpendicular component.  $p$ is electron momentum, $v$ is the velocity, and $\theta$ is the pitch angle. $v_{\parallel}=v\cos\theta$, $v_{\perp}=v\sin\theta$. $\omega_{cb}=\omega_{ce}/\gamma$ is the cyclotron frequency for a relativistic electron ($\gamma$ is the relativistic factor). $f$ is the electron distribution function normalized to $\int d^{3}\mathbf{p} f=1$, and $J_{n}$ is the $n$th order Bessel function, with argument $k_{\perp} v_{\perp}/\omega_{cb}$.
\endgroup

According to Eqs. (\ref{eq:epsilonA}) and (\ref{eq:K}), both the wave emission and absorption happens with the electrons satisfying the resonance condition with the wave,
\begin{equation}
\label{eq:resonance-condition}
  \omega-k_{\parallel}v_{\parallel}-n\omega_{cb}=0.
\end{equation}
For unrelativistic electrons $k_{\parallel}=0$, the resonance condition can be simplified as $\omega=n\omega_{ce}$ with $n<0$. The wave-particle
resonance only happens in the regions where the value of $B$ satisfies the resonance condition, which is called the resonance region. Regarding the absorption efficiency in the resonance region, the plasma can be categorized as optically thick or optically thin for different waves. In the optically thick case, the wave will be strongly damped in the resonance region (the characteristic length of damping $L=1/k_{i}$ is much smaller than the scale length of the resonance region), which can be regarded as a black body with radiation power only depending on the temperature. In the optically thin case, the wave will be weakly absorbed in the resonance region, and the radiation power depends on density and length of the resonance region as well\cite{tobias_ece-imaging_2012,shi_synthetic_2016}. Typically for X-wave with $n\ge -2$, the plasma can be treated as optically thick, and for O-wave or smaller $n$ the plasma is optically thin.

For suprathermal electrons like REs, the wave-particle resonance can happen in regions
outside the resonance regions due to the relativistic factor. These electrons
can provide emission and absorption in addition to the thermal radiation. However, the emission of ECWs also depends on the pitch angle of the resonant electrons, through the Bessel function $J_{n}$ in Eq. (\ref{eq:K}). The pitch angle requirement is particularly important for high-energy electrons and high frequency ECWs, where $|n|$ is large, and the Bessel function $J_{n}$ at higher $|n|$ requires larger argument (comparable to $|n|$) to get a significant output value. This condition can limit the ECE from the runaway electron tail, since the pitch angle distribution of RE is usually strongly collimated due to the weak collisional scattering in the high-energy regime.

On the other hand, if the suprathermal electrons can have have large pitch angle, then the ECE from them will differentiate from thermal electrons not only in frequencies and resonances, but also in polarizations. According to the polarization of X-wave and O-wave, we know that the radiation of X-wave depends on the $xx$, $yy$ and $xy$ component of $\underline{\underline{K}}$, whereas the radiation of O-wave depends on the $zz$ component. Given Eq. (\ref{eq:Tn}) and taking the leading order term of the Bessel functions, we find that the X-wave emission is proportional to $\left(k_{\perp} v_{\perp}/\omega_{cb}\right)^{2n-2}$, while the O-wave emission is proportional to $\left(k_{\perp} v_{\perp}/\omega_{cb}\right)^{2n}$. For thermal electrons, it is known that $k_{\perp} v_{\perp}/\omega_{cb}\ll 1$, thus the X-wave emission dominates the O-wave. However, for superthermal electrons with large pitch angle, $v_{\perp}\sim c$, we find that the O-wave emission can be comparable to X-wave, which leads to an ECE with a combined polarization.

According to Eqs. (\ref{eq:epsilonA}) and (\ref{eq:K}), for a Maxwellian electron distribution with temperature $T_{e}$, $\underline{\underline{\epsilon}}^{Ar}$ and $\underline{\underline{K}}$ satisfy
\begin{equation}
\label{eq:fluctuation-dissipation}
  \frac{K_{ij}}{\epsilon^{Ar}_{ij}}=\frac{\omega T}{\pi}
\end{equation}
for every term in the tensor. This result is an outcome of the fluctuation-dissipation theorem, which states that for a system in thermal equilibrium, the fluctuation amplitude level is always proportional to the efficiency of the corresponding dissipative process, with the ratio proportional to the system temperature.

We can also define the normalized emission term,
\begin{equation}
\label{eq:U}
  \textrm{For X-wave,}\qquad U=\frac{\pi\omega}{c^{2}}\frac{\mathbf{E}^{\dagger}\cdot \underline{\underline{K}}\cdot\mathbf{E}}{2 k |E_{y}|^{2}}
\end{equation}
\begin{equation}
\label{eq:U}
\textrm{For O-wave,}\qquad U=\frac{\pi\omega}{c^{2}}\frac{\mathbf{E}^{\dagger}\cdot \underline{\underline{K}}\cdot\mathbf{E}}{2 k |E_{z}|^{2}}
\end{equation}
If we only consider $k_{i}$ from the resonance absorption $\underline{\underline{\epsilon}}^{Ar}$,  we find $U/k_{i}=T_{e}$ according to Eq. (\ref{eq:ki})(\ref{eq:fluctuation-dissipation}). The radiation power is then
\begin{equation}
\label{eq:reciprocal2}
 P=\frac{c^{2}}{\pi\omega}\int dx 2U(x)  \exp\left(-2\int dx k_{i}(x)\right).
\end{equation}

Assuming all the power of the artificial wave is absorbed by the plasma with uniform temperature $T_{e}$, the ECE radiation power can be calculated as
\begin{equation}
  P=\frac{c^{2}T_{e}}{\pi\omega}\int dx 2k_{i} \exp\left(-2\int dx k_{i}\right)=\frac{c^{2}T_{e}}{\pi\omega}
\end{equation}
thus the radiation power is proportional to the plasma temperature. This result is consistent with Kirchhoff's radiation law. In the later text, we will use the effective radiation temperature $T_{\mathbf{eff}}=(\pi\omega/c^{2})P$ to represent the ECE radiation power, which characterizes the plasma temperature in the resonance region for the optically thick case. 

In addition to resonance absorption, the collision effect ($\underline{\underline{\epsilon}}^{Ac}$) can also contribute to the wave damping and increase $k_{i}$. This will results in $U/k_{i}<T_{e}$ and the final radiation power will be smaller than that from Kirchhoff's radiation law. In the normal tokamak experimental parameters like what will be discussed in Sec. \ref{sec:ECE-simulations}, where $T_{e}\sim $keV, the collisional damping is much smaller than the resonance absorption and can be ignored. However, for post-disruption scenarios where $T_{e}\sim$ a few eV, the resonance absorption from thermal electrons is very weak and the collisional damping is significantly enhanced. In this case, the contribution from collisions dominates the damping of ECWs.

In the numerical model, the integrals in Eq. (\ref{eq:reciprocal2}) are calculated using the  finite difference method. For each reciprocal calculation, 10 reflections are accounted in total, with $\alpha_r=0.76$ and $\alpha_p=0.20$. The profile of the magnetic field $B$ is set as $B(x)=B_{0}R/x$ where $R$ is the major radius and $B_{0}$ is the magnetic field on the axis. In calculating $k_{i}$ and $U$ at every positions, the resonance harmonics $n$ are calculated from $-2$ to $-20n_{s}$, where $n_{s}=\omega/\omega_{ce0}$ and $\omega_{ce0}=eB_{0}/mc$. The electron distribution function $f$ is obtained from a kinetic simulation, which is discussed in Sec. \ref{sec:RE-simulation}.

\section{Simulation of RE momentum space distribution including kinetic instabilities}
\label{sec:RE-simulation}

In order to model the non-thermal ECE radiation, the electron distribution function in momentum space needs to be modeled correctly. Especially for runaway electrons, the distribution function has significant deviations from a Maxwellian and cannot be simply modeled as a bi-Maxwellian. Here the distribution function is calculated by solving the spatially homogeneous kinetic equation. The coordinates for momentum space are $(p,\xi)$, where $p$ is the electron momentum normalized to $mc$ ($m$ is the electron mass and $c$ is the speed of light), and $\xi=p_{\parallel}/p$ is the cosine of the pitch angle. For runaway electrons, the important forces affecting momentum space dynamics are the electric force, the collisions, and the radiation reaction forces. Knock-on collisions which are ignored in the Fokker-Planck form of the collision operator provide an additional source for RE generation. The kinetic equation we solve can be written as
\begin{equation}
\label{eq:kinetic-equation}
\frac{\partial f}{\partial t}+\frac{eE_{\parallel}}{mc}\left(\xi \frac{\partial f}{\partial p}+\frac{1-\xi^2}{p}\frac{\partial f}{\partial \xi}\right)+C\left[f\right]+\frac{\partial}{\partial \mathbf{p}}\cdot\left(\mathbf{F}_{\mathrm{rad}}f\right)+D\left[f\right]=S_{A}\left[f\right],
\end{equation}
where $E_{\parallel}$ is the parallel electric field and $e$ is the electron charge, $C[\dots]$ represents the test particle collision operator\cite{landreman_numerical_2014}, $\mathbf{F}_{\mathrm{rad}}$ is the synchrotron radiation reaction force\cite{stahl_effective_2015}, $D[\dots]$ is the quasilinear diffusion operator from the excited waves, and $S_{A}[\dots]$ is the source term for the knock-on collisions.

\begin{figure}[h]
	\begin{center}
		\includegraphics[width=0.9\linewidth]{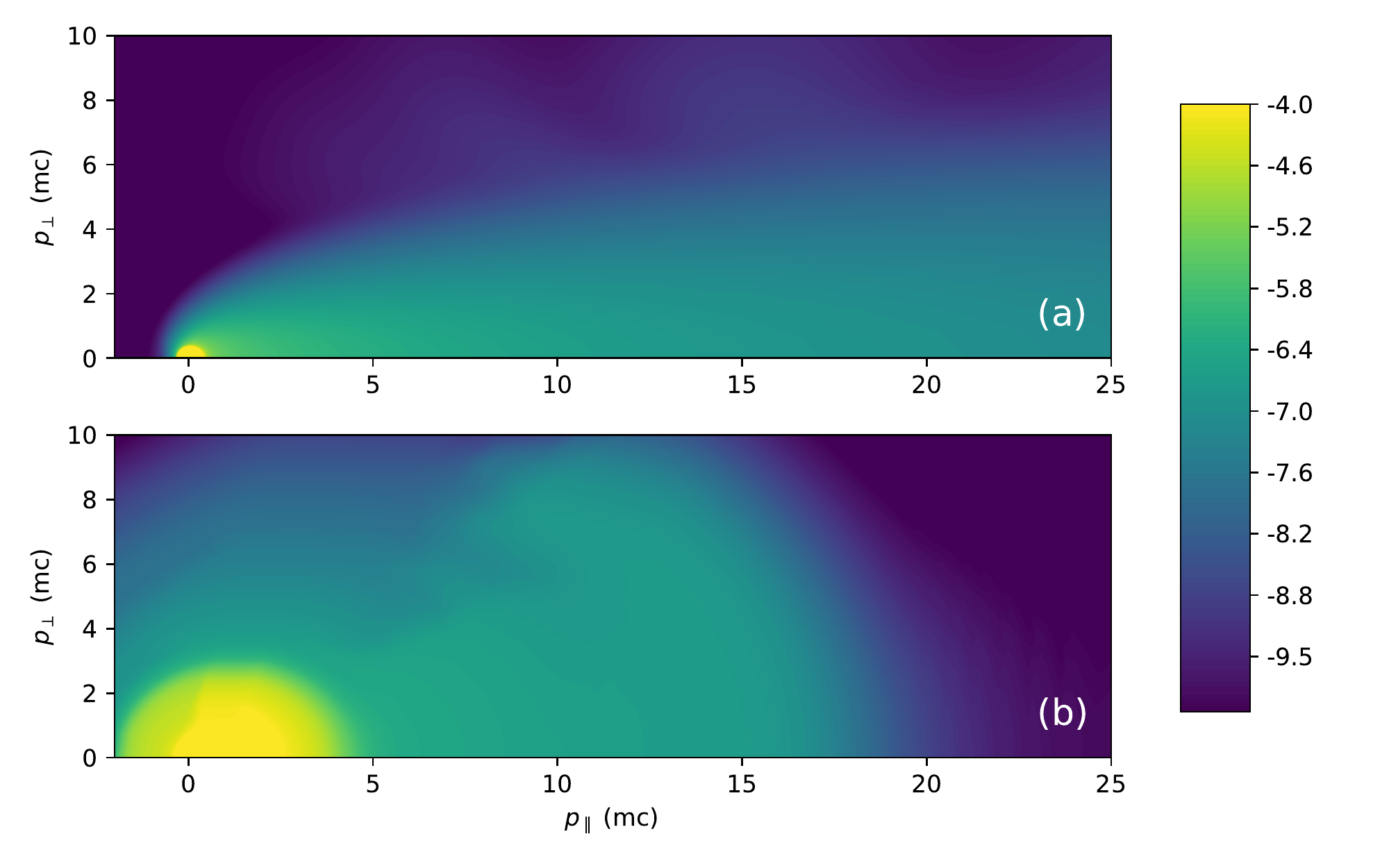}
	\end{center}
	\caption{\label{fig:f2d} RE distribution function in 2D momentum space from kinetic simulations, with (a) kinetic instabilities turned off and (b) kinetic instabilities turned on. }
\end{figure}

In this model we take into account the kinetic instabilities associated with the RE beam, and its back-reaction on the RE distribution function using the quasilinear diffusion model in magnetized plasmas\cite{kaufman_quasilinear_1972}. The RE distribution determined by all the kinetic forces has a wide distribution in energy, but is strongly forward beamed and thus has a very narrow distribution in pitch angle (see Fig. \ref{fig:f2d} (a)). This anisotropic distribution can be susceptible to various kinds of kinetic instabilities, most notably one driven by its interaction with the whistler waves\cite{fulop_destabilization_2006}. In addition, a bump-on-tail distribution formed by the energy dissipation from radiation reaction\cite{hirvijoki_radiation_2015} can also driven plasma waves. The excited plasma waves can scatter the resonant electrons strongly in pitch angle when satisfying the resonance condition (Eq. (\ref{eq:resonance-condition})), and isotropize the distribution, called "fan instabilities"\cite{parail_kinetic_1978}. Given that the ECE depends sensitively on the pitch angle of electrons, these kinetic instabilities can have a remarkable impact on the ECE radiation.

\begin{figure}[h]
	\begin{center}
		\includegraphics[width=0.5\linewidth]{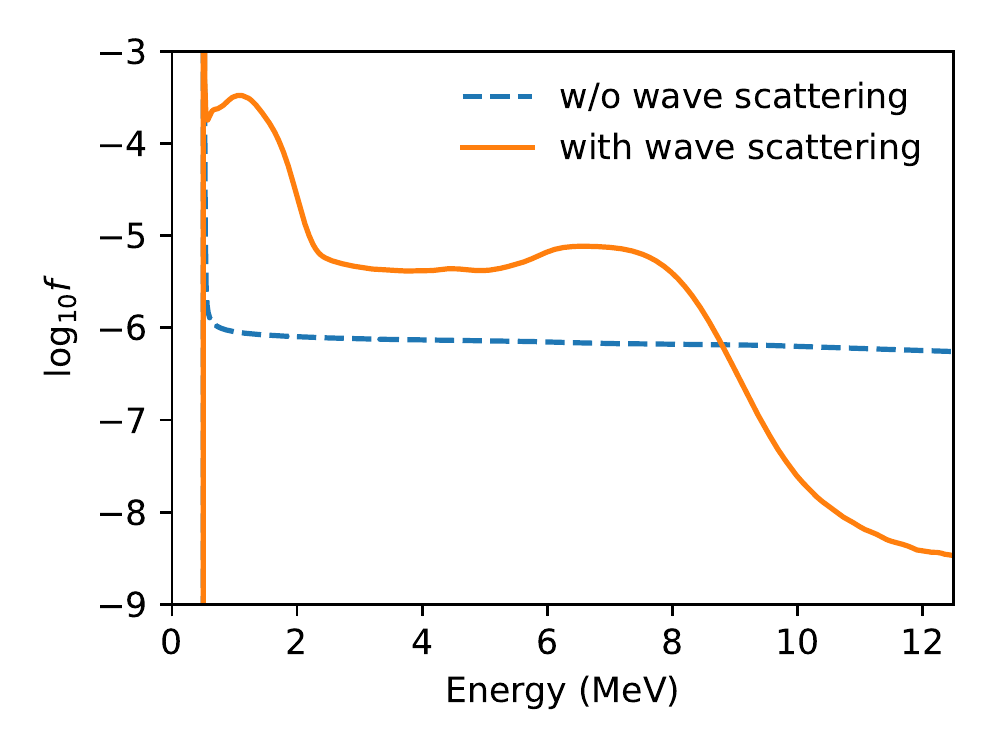}
	\end{center}
	\caption{\label{fig:f1d} RE distribution function in energy. Dashed line is the result from simulation with kinetic instabilities turned off, and solid line is the result with kinetic instabilities turned on. }
\end{figure}

The impact of kinetic instabilities on the RE distribution is illustrated in Fig. \ref{fig:f2d} and Fig. \ref{fig:f1d}, where we show a comparison of the RE distribution functions obtained from the kinetic simulation, with quasilinear diffusion operator turned on and off. The plasma parameters are the same as the simulation in Sec. \ref{sec:ECE-simulations} from a DIII-D flattop experiment. As shown in Fig. \ref{fig:f2d}, the excited modes lead to an increase of RE pitch angle in both low-energy and high-energy regimes. This scattered electron distribution affected by kinetic instabilities (Fig. \ref{fig:f2d} (b)) is quantitatively different from the unscattered distribution (Fig. \ref{fig:f2d} (a)).  Especially in the low-energy regime, where the resonant electrons can be scattered to $p_{\parallel}\le0$, which can significantly enhance the emission of ECWs. In addition to scattering, interactions with the whistler waves also cause a higher avalanche growth rate and a larger RE population, as shown in Fig. \ref{fig:f1d}. Affected by the waves, the electrons tend to accumulate in low-energy regime and stop going to very high energy. The value of the critical electric field is also altered by the kinetic instabilities\cite{liu_role_2018}.


Knock-on collisions between highly energetic REs and low-energy thermal electrons cause an exponential growth of the RE population, called "runaway electron avalanche". This effect can be accounted for using a source term in the kinetic equation derived from the Boltzmann collision operator using a M{\o}ller cross section\cite{moller_zur_1932}. In previous studies of the source term\cite{rosenbluth_theory_1997,chiu_fokker-planck_1998}, the pitch angle distribution of the seed REs is ignored based on the assumption that the RE distribution is strongly anisotropic. Given the new distribution from the simulation including kinetic instabilities, this assumption is not valid. In terms of this, we apply a new knock-on collision source term\cite{boozer_theory_2015,liu_adjoint_2017} in Eq. (\ref{eq:kinetic-equation}) which takes into account both the energy and the pitch angle distribution of the seed REs. This source term is calculated by transforming  the distribution function to a spectrum representation using Legendre polynomials\cite{landreman_numerical_2014}, in which the knock-on collision operator can be simplified.

Note that in the current model the distribution function is only solved in momentum space. In order to do a synthetic diagnostic of ECE, we need the electron distribution functions along the ray path. Based on the knowledge of RE generation and current diagnostic results\cite{paz-soldan_spatiotemporal_2017}, it is known that most of the generated runaway electrons live in the region close to the core. Thus for simplicity, we choose a tokamak cross-section with $r<0.5a$ as the RE region, where $a$ is the minor radius. Inside this region, $f$ is homogeneous in space, initialized as a Maxwellian with $T_{e}=T_{e0}$, and evolves with the kinetic simulation. Outside this region, $f$ is assumed to be a Maxwellian with temperature profile $T_{e}=(T_{\mathrm{core}}-T_{\mathrm{edge}})(1-(r/a)^{2})^{2}+T_{\mathrm{edge}}$, and unchanged with time. Plasma density is assumed to have a similar profile $n_{e}=n_{e0}(1-(r/a)^{2})^{2}$.

\section{ECE signals from runaway electron simulations}
\label{sec:ECE-simulations}

We now use the ECE synthetic diagnostic model to calculate the ECE signals from the RE simulations. The simulation is set based on the low-density flattop RE experiments on DIII-D\cite{paz-soldan_non-thermal_2016,paz-soldan_spatiotemporal_2017}. The experimental discharge contains two stages. In the first stage, the plasma density is low and the runaway electron tail is driven by a parallel electric field from the Ohmic coils. In the second stage, the plasma density increases due to gas puffing and the value of $E/E_{CH}$ decreases ($E_{CH}$ is the Connor-Hastie critical electric field\cite{connor_relativistic_1975}). An experimental example of the two stages divided by gas puffing can be found in \cite{paz-soldan_spatiotemporal_2017}. The parameters we used are close to the numbers from the tokamak core diagnostic. For stage 1, density $n_{e0}$ is $0.6\times 10^{19}$m$^{-3}$, temperature $T_{e0}$ is $1.3$keV, $T_{\mathrm{core}}$ is $2.0$keV and $T_{\mathrm{edge}}$ is $0.2$keV. The electron distribution $f$ is initialized as a Maxwellian. Electric field $E_{\parallel}=0.055$V/m. For stage 2, $n_{e0}$ is $0.8\times 10^{19}$m$^{-3}$. The temperature profile is the same as stage 1 and the electron distribution is initialized from the final step of stage 1. The electric field $E_{\parallel}$ is $0.045$V/m. For the dimension of the device, the major radius $R$ is $2.0$m and the minor radius $a$ is $0.5$m. The magnetic field on axis in two stages is $B_{0}=1.45$T.

\begin{figure}[h]
	\begin{center}
		\includegraphics[width=0.9\linewidth]{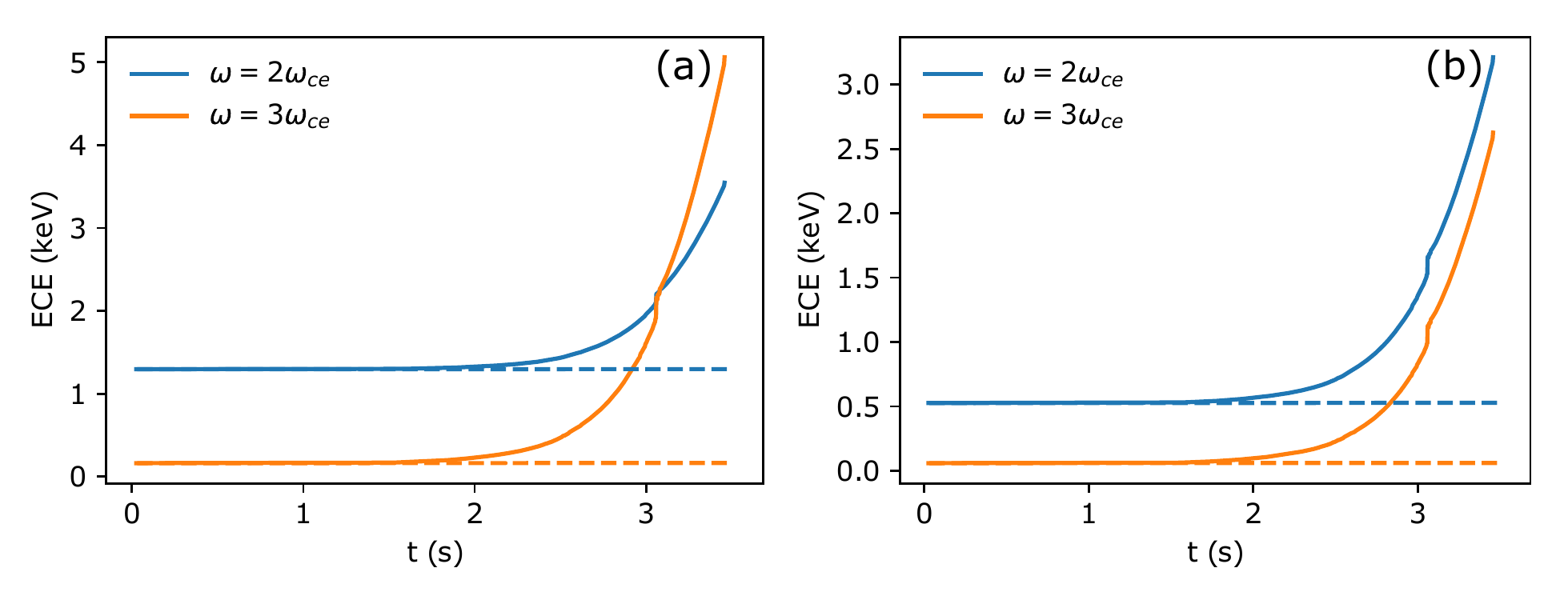}
	\end{center}
	\caption{\label{fig:ece1}ECE signals from RE simulation stage 1, for (a) X-mode and (b) O-mode. Solid lines are results with kinetic instabilities and wave-particle interactions, and dashed lines are results with kinetc instabilities turned off.}
\end{figure}

The results are summarized as follows. In Fig. \ref{fig:ece1} we show the effective radiation temperature $T_{\mathrm{eff}}$ of ECE signals calculated from the synthetic diagnostic, for two different polarizations and at two different frequencies in stage 1. The frequencies of ECE are chosen to be $2\omega_{ce0}$ and $3\omega_{ce0}$. The solid lines are calculated from the scattered RE distribution including the wave-particle interactions (Fig. \ref{fig:f2d} (b)). The dashed lines are the results from the unscattered RE distribution function (Fig. \ref{fig:f2d} (a)).

At the beginning of stage 1, the value of the $T_{\mathrm{eff}}$ for the X-mode with $\omega=2\omega_{ce0}$ is very close to $T_{e0}$, which is consistent with Kirchhoff's law. The value of the ECE signal for $\omega=3\omega_{ce0}$ is much smaller, since the plasma is optically thin for this high frequency wave. The O-mode signals mostly come from the polarization scrambling, given its own radiation is small compared to X-mode. During the RE population growth in stage 1, the ECE signals barely grow until 3.0s. After 3.0s, the ECE signals for both polarizations from the scattered RE distribution grow abruptly, whereas the ECE signals from the unscattered distribution remain dormant. An interesting feature of ECE signals in this abrupt growing phase is that for the X-mode, the $\omega=3\omega_{ce0}$ signal surpasses the $\omega=2\omega_{ce0}$ signal. Comparing with the experiments (Fig. 5 (d) in \cite{paz-soldan_non-thermal_2016}), we find that the X-mode ECE results with the scattered distribution has much better agreement with the experiments, including the growth rate, peak amplitudes and the overpassing behavior. This is strong evidence that the kinetic instabilities are present and play an important role in the DIII-D QRE experiments\cite{liu_role_2018}.

\begin{figure}[h]
	\begin{center}
		\includegraphics[width=0.9\linewidth]{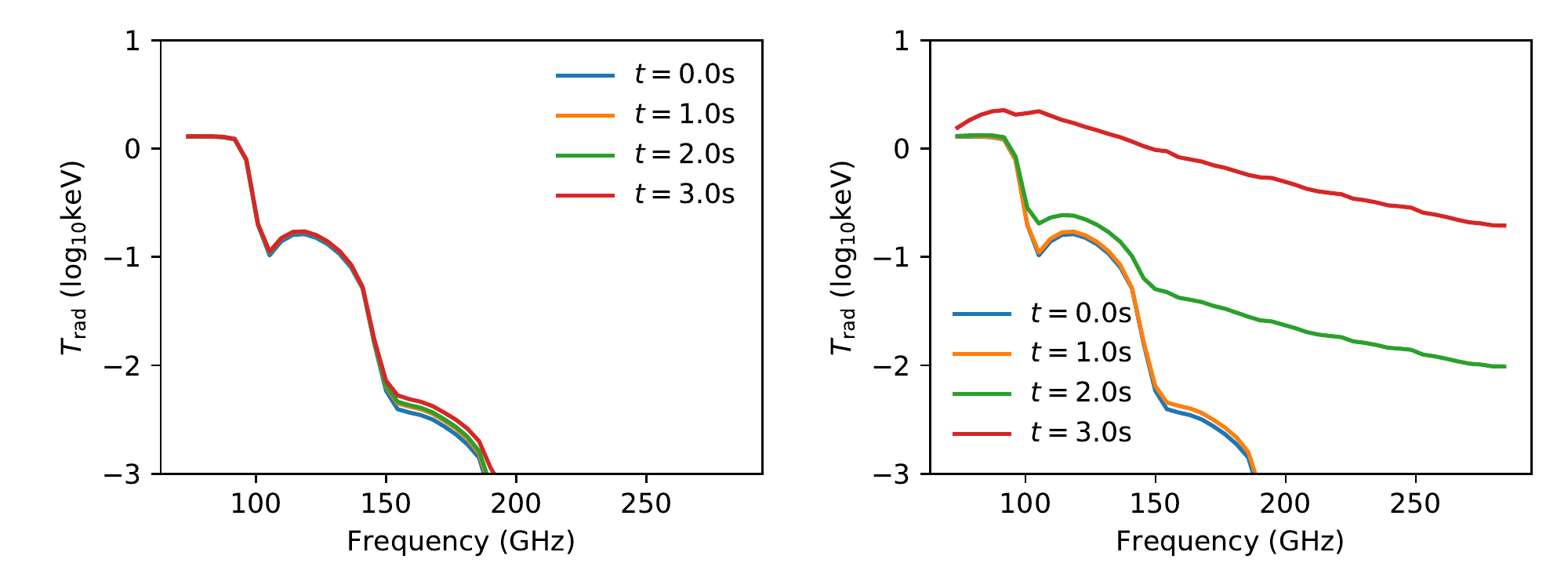}
	\end{center}
	\caption{\label{fig:ece-spectrum}Evolution of X-mode ECE signal spectrum in stage 1, for (a) simulation with kinetic instabilities turned off, and (b) simulation with kinetic instabilities present.}
\end{figure}

\begin{figure}[h]
	\begin{center}
		\includegraphics[width=0.5\linewidth]{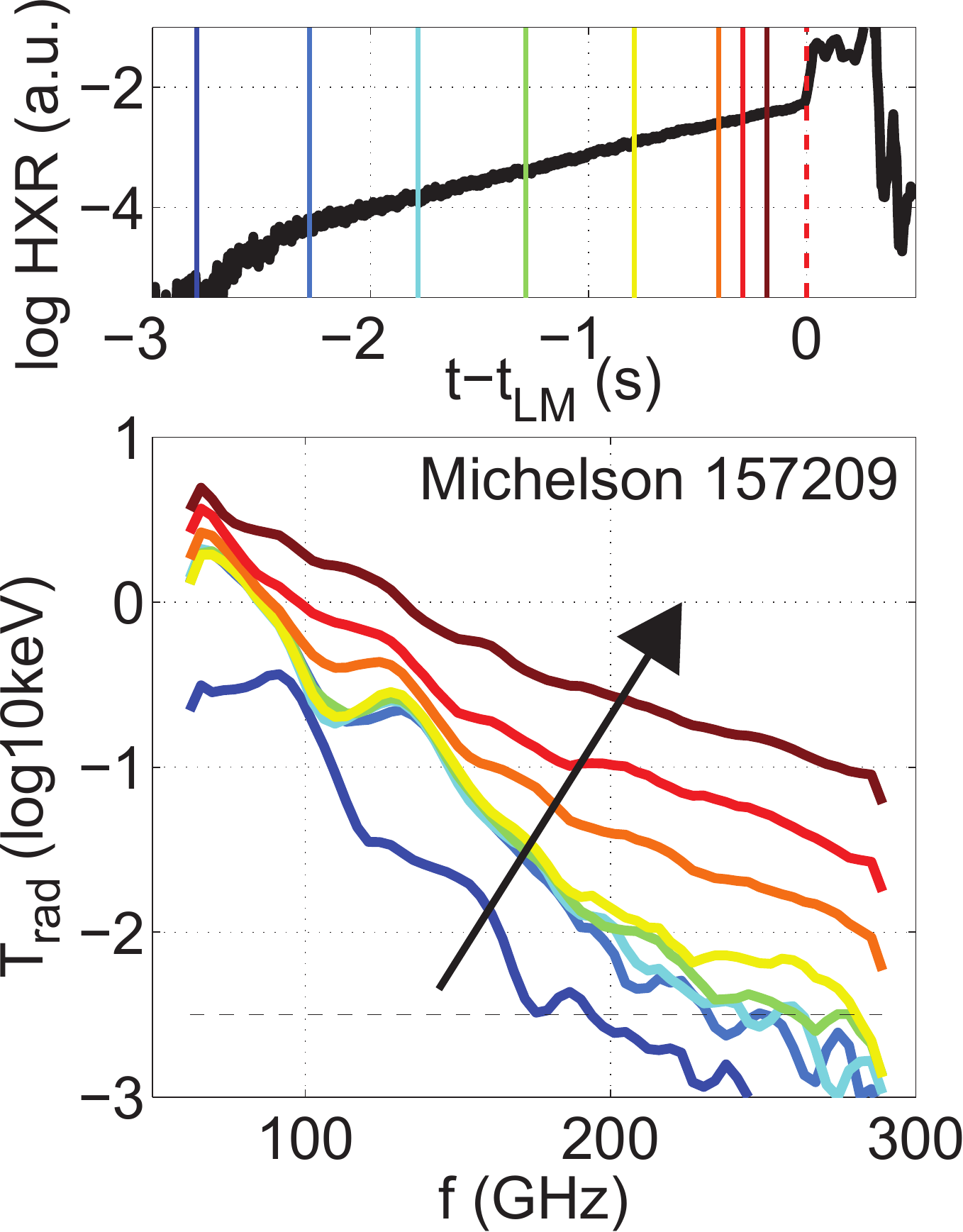}
	\end{center}
	\caption{\label{fig:ece-spectrum2}The ECE spectrum evolution measured from Michelson interferometer in DIII-D shot 157209. The lines correspond to different time during the RE population growth, as indicated by the vertical lines in the Hard X-ray plot.}
\end{figure}

The different behaviors of the ECE signals at two frequencies inspire us to study the change of the whole ECE spectrum during RE growth. In Fig. \ref{fig:ece-spectrum} we show the spectrum of the X-mode ECE signals at different times in stage 1, for the unscattered and the scattered distributions. At the beginning when $f$ is close to a Maxwellian, the spectrum is like a step function, which reflects the uniform radial profile of the RE distribution function. At a later time, the spectrum for the unscattered distribution changes little, whereas for the scattered distribution the ECE signals grow at all frequencies and the spectrum becomes flatter, meaning that REs give larger contributions to ECE at higher frequency ECE than the lower frequencies. This flattening behavior is consistent with the ECE spectrum observed in experiments\cite{paz-soldan_growth_2014} (Fig. \ref{fig:ece-spectrum2}).

\begin{figure}[h]
	\begin{center}
		\includegraphics[width=0.5\linewidth]{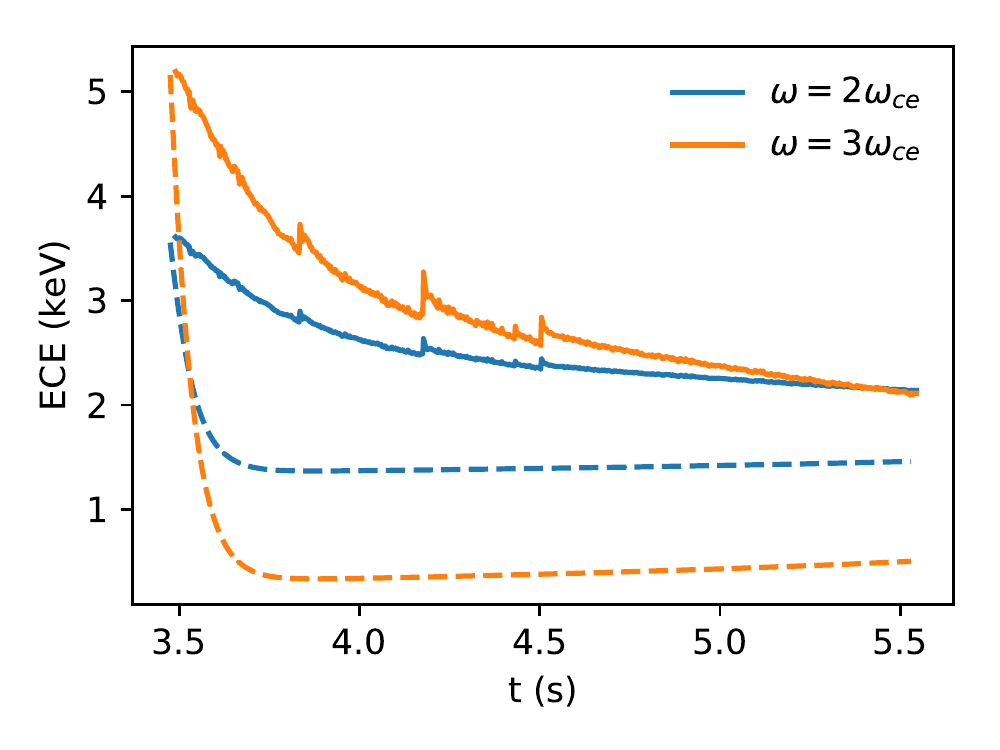}
	\end{center}
	\caption{\label{fig:ece2}X-mode ECE signals from RE simulation stage 2. Solid lines are results with kinetic instabilities, and dashed lines are results without these effects.}
\end{figure}

Fig. \ref{fig:ece2} shows the X-mode ECE signals at two frequencies in stage 2, with and without the effects of kinetic instabilities. The initial states of $f$ are both from the scattered distribution functions in stage 1. When the kinetic instabilities are turned off, the ECE signals drop to its base values very quickly (in 0.2s), and then stay at these low levels, despite the fact that there remains a large population of high-energy REs. One the other hand, with the wave diffusion turned on, the ECE signals stay at a high level and drop gradually, which is consistent with the evolution of RE density and agrees with experiments. This further confirms that the high level of non-thermal ECE signals observed is a reflection of not only the large population of runaway electrons generated, but also the intensity of the excited kinetic instabilities.

\begin{figure}[h]
	\begin{center}
		\includegraphics[width=0.5\linewidth]{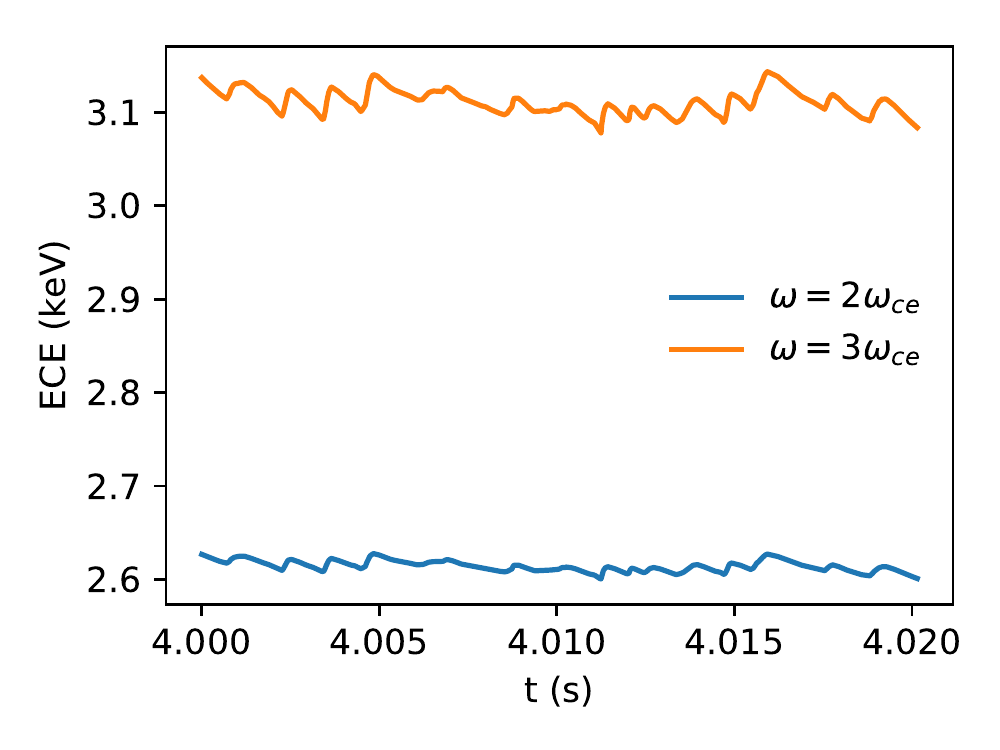}
	\end{center}
	\caption{\label{fig:ece3}Oscillations of X-mode ECE signals in stage 2.}
\end{figure}

In addition to the trends of ECE signals within the discharge, we also study the behavior of the ECE signals on short timescales. Examination of the simulation results shows that the X-mode ECE signals in stage 2 has oscillatory behaviors, as shown in Fig. \ref{fig:ece3}, with periods of about 0.002s. In each period, the ECE signals experiences a fast growth phase, followed by a slower decaying phase. This behavior is consistent with the "inverse-sawtooth" behavior of ECE signals observed in QRE experiments\cite{spong_first_nodate,zhou_investigation_2013} and in post-disruption scenarios\cite{fredrickson_control_2015}.

\section{Explanations of simulation results}
\label{sec:explanation}

\begin{figure}[h]
	\begin{center}
		\includegraphics[width=0.9\linewidth]{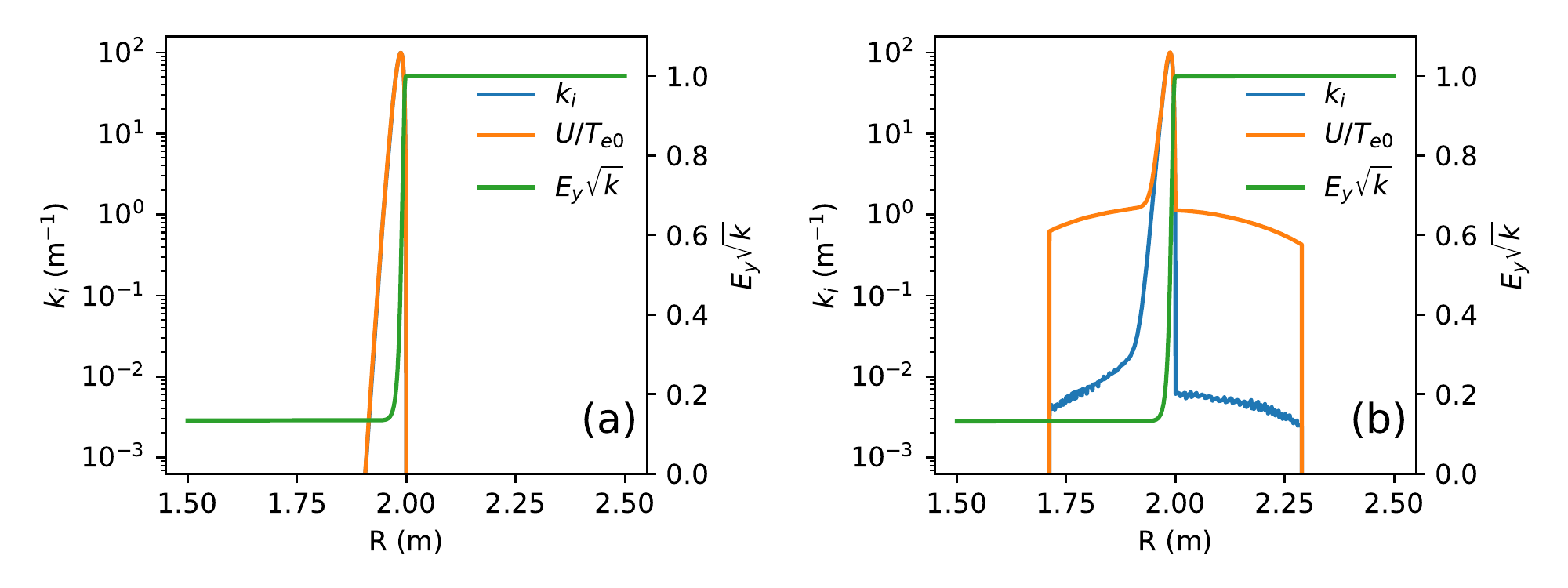}
	\end{center}
	\caption{\label{fig:ece-contrib1}Normalized emission coefficient $U/T_{e0}$ (Eq. (\ref{eq:U})), absorption coefficient $k_{i}$ (Eq. (\ref{eq:ki})), and the amplitude $E_{y}$ of the artificial X-mode electron cyclotron wave with $\omega=3\omega_{ce0}$, calculated from (a) a Maxwellian distribution, and (b) RE tail as shown in Fig. \ref{fig:f2d} (b).}
\end{figure}

To have a better understanding of the results obtained in Sec. \ref{sec:ECE-simulations}, here we show the details of the calculation using the reciprocity method.
We first look at the results of the emission and absorption profiles. Fig. \ref{fig:ece-contrib1} shows the profiles of $k_{i}$ (Eq. (\ref{eq:ki})) and $U/T_{e0}$ (Eq. (\ref{eq:U})) in the plasma (from $R=1.5$m to 2.5m) in stage 1, for X-mode ECWs with $\omega=2\omega_{ce0}$. At the beginning of stage 1 when the electron distribution is close to a Maxwellian (Fig. \ref{fig:ece-contrib1} (a)), both the emission and the absorption are strongly localized near the magnetic axis, where the resonance region for this wave is located. The value of $U/T_{e0}$ is almost equal to $k_{i}$ in the resonance region, consistent with Eq. (\ref{eq:fluctuation-dissipation}).

Fig. \ref{fig:ece-contrib1} (b) shows the emission and absorption profiles after a significant population of RE generated, including effects of wave-particle interactions (Fig. \ref{fig:f2d} (b)). We can see that the RE tail increases the amplitudes of emission and absorption outside the resonance region, as discussed in Sec. \ref{sec:ECW}. As shown by the orange and blue lines outside the resonance region, the ratio between the emission and absorption from RE tail is not equal to $T_{e0}$ but much larger. This means that REs contribute to strong emission of ECWs but weaker absorption, compared to thermal electrons. This difference originates from the shapes of distribution function $f$. Note that in Eqs. (\ref{eq:K}) and (\ref{eq:epsilonA}), the emission coefficient depends on the value of $f$ satisfying the resonance condition, whereas the absorption coefficient depends on the gradients of $f$. As shown Fig. \ref{fig:f1d} and Fig. \ref{fig:f2d}, the RE tail distribution has a much smoother shape compared to the Maxwellian part, resulting in a weaker absorption for the ECWs. The wave-particle interaction can further broaden the difference by making the RE tail distribution smoother.
The result of artificial wave calculation, which only depends on the absorption, is almost identical in the two cases, as shown by the green line in Fig. \ref{fig:ece-contrib1}.

\begin{figure}[h]
	\begin{center}
		\includegraphics[width=0.9\linewidth]{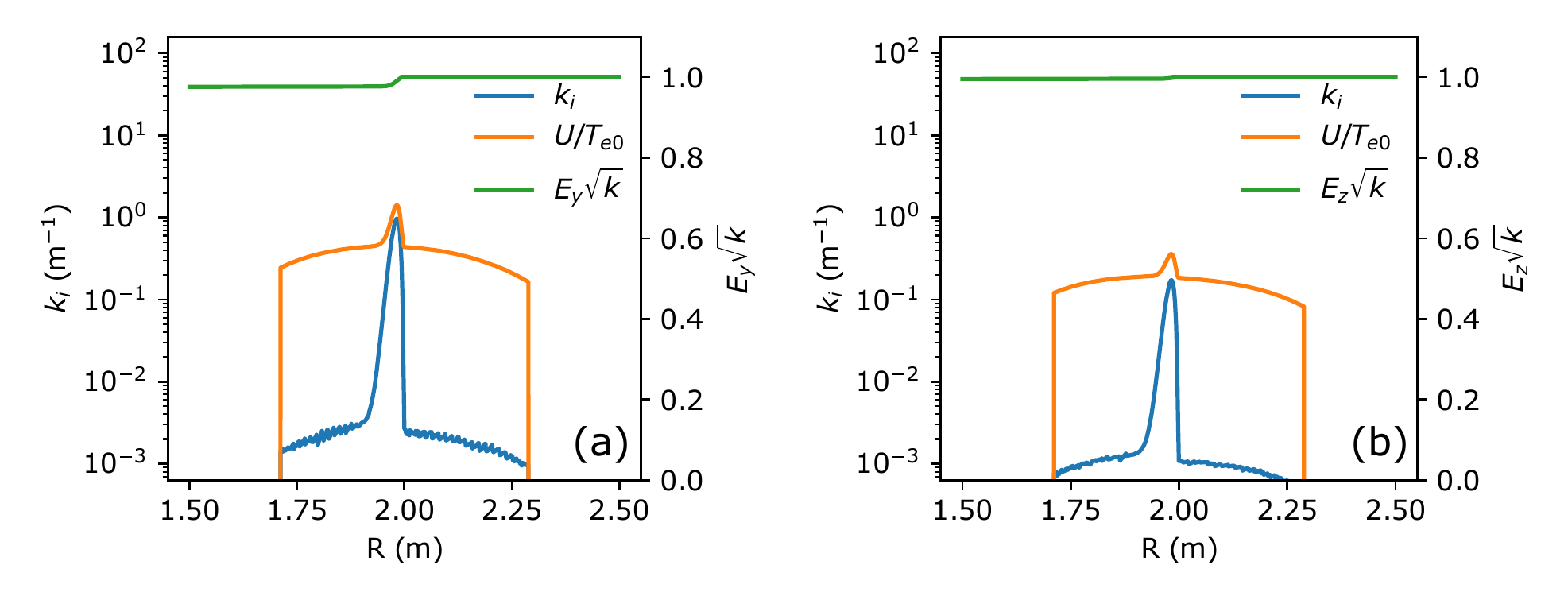}
	\end{center}
	\caption{\label{fig:ece-contrib2}Normalized emission coefficient $U/T_{e0}$, absorption coefficient $k_{i}$, and the amplitude $E_{y}$ of the artificial  electron cyclotron wave for (a) X-mode with $\omega=3\omega_{ce0}$ and (b) O-mode with $\omega=2\omega_{ce0}$. Both results are calculated for a RE tail distribution in Fig. \ref{fig:f2d} (b).}
\end{figure}

Fig. \ref{fig:ece-contrib2} shows the emission and absorption profiles for X-mode ECW with $\omega=3\omega_{ce0}$, and the O-mode ECWs with $\omega=2\omega_{ce0}$, both calculated from the RE tail distribution. We can see that for ECWs with higher frequency or different polarization, both the absorption and the emission in the resonance region is much weaker compared to Fig. \ref{fig:ece-contrib1}. This leads to a difference in the optical thickness. As shown by the green lines in Fig. \ref{fig:ece-contrib1} and Fig. \ref{fig:ece-contrib2}, the damping of the artificial waves are very different in the resonance regions.
Unlike the thermal electrons, the level of ECE from runaway electrons outside resonance region (the orange lines) at different  frequencies and polarization are close, meaning that the RE tail takes a more important role in the ECE radiation power. This explains that the ECE spectrum will become flatter when the contributions from RE dominate. It also illustrate that the growth of O-wave ECE found in Sec. \ref{sec:ECE-simulations} is not only due to the polarization scrambling at walls, but also comes from the enhanced emission of O-wave from the runaway electrons.
The weak damping of $\omega=3\omega_{ce0}$ wave and the O-wave means that the plasma cannot be considered as a black body and the received ECE power will thus be smaller than $T_{e0}$ even for a Maxwellian distribution. In addition, it also means that the emission of ECWs from REs in the whole domain can propagate to the receiver, whereas the for the X-wave with $\omega=2\omega_{ce0}$, the emission from the low-field-side will be largely absorbed in the resonance region. Thus the ECE power for $\omega=3\omega_{ce0}$ can be larger than $2\omega_{ce}$ if the emission from RE dominates.

To illustrate the dependence of the ECE power on runaway electrons at different momentum, we introduce the weight function for ECE radiation. Note that in Eq. (\ref{eq:reciprocal2}), the absorption of the artificial wave $k_{i}$ mainly depends on the thermal electrons and the wall reflection, since the absorptions by REs are weak. The emission term $U$ depends on the both the Maxwellian part and the RE tail through Eqs. (\ref{eq:K}) and (\ref{eq:U}). In terms of that, we can substitute a delta function $\delta(\mathbf{p}-\mathbf{p}_{0})$ for $f$ into Eq. (\ref{eq:K}) to obtain $U$, and  calculate the ECE weight function $W(\mathbf{p}_{0})$ with $k_{i}$ from a Maxwellian distribution. The radiation effective temperature can then be expressed as
\begin{equation}
  T_{\mathrm{eff}}=\int d^{3}\mathbf{p}_{0} W(\mathbf{p}_{0}) f(\mathbf{p}_{0}).
\end{equation}

\begin{figure}[h]
	\begin{center}
		\includegraphics[width=0.9\linewidth]{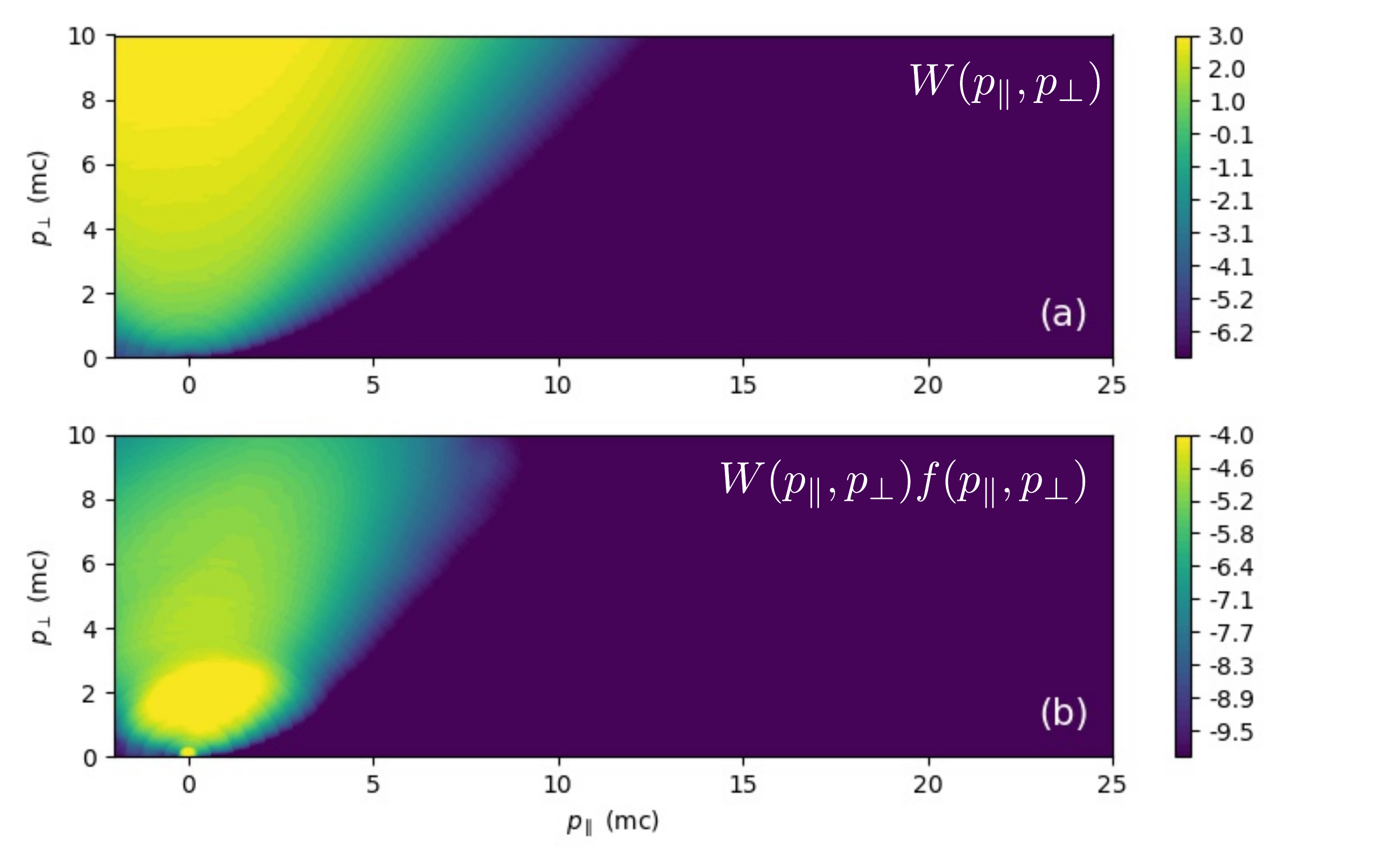}
	\end{center}
	\caption{\label{fig:weight} (a) Weight function of ECE signals $W(p_{\parallel}, p_{\perp})$ in $\log_{10}$ scale in momentum space for $\omega=3\omega_{ce0}$. (b) Result of $W(p_{\parallel}, p_{\perp})f(p_{\parallel},p_{\perp})$ in $\log_{10}$ scale for the scattered distribution. }
\end{figure}

Fig. \ref{fig:weight} shows the values of $W(\mathbf{p}_{0})$ and $W_{\mathbf{p}_{0}}f(\mathbf{p}_{0})$ in RE momentum space, for X-mode ECWs with $\omega=3\omega_{ce0}$ and a scattered runaway electron distribution. We can see that only runaway electrons with large pitch angles can give significant contributions to ECE power. As shown in Fig. \ref{fig:weight} (b), for the scattered RE distribution the ECE power comes from both thermal electrons (the small circular region near $p_{\parallel}=p_{\perp}=0$) and the runaway electrons with $p<5$ and pitch angle larger than $\pi/6$. The low-energy runaway electrons with large pitch angles are a result of the wave-particle interaction from the high frequency whistler waves\cite{liu_role_2018}. This explains why the unscattered RE distribution gives little ECE radiation power, since they have little population in this region (Fig. \ref{fig:f2d} (a)).

According to the ECE weight function, the ECE signals mainly characterize the RE density in the low-energy regime ($<2.5$MeV), which is different from the SE and gamma ray radiation which mainly comes from very high-energy runaway electrons. The ECE power is also very sensitive to the pitch angle distribution. Given the shape of the weight function, a high level of ECE from RE can be regarded as a signal of anomalous pitch angle scattering of REs in the low-energy regime, and an indication of RE kinetic instabilities.

The oscillatory behavior of the ECE signals shown in Fig. \ref{fig:ece3} is related to the nonlinear dynamics of the wave-particle interactions. This behavior has been observed in DIII-D QRE experiments, and is found to be correlated with the excitation and damping of whistler waves with frequency $100-200$MHz\cite{spong_first_nodate}. As shown by the weight function in Fig. \ref{fig:weight}, the high level ECE signals are sensitive to the population of REs with large pitch angles. Given that the excitation of whistler waves requires a strong gradient of $f$, and the quasilinear diffusion tends to  smoothen the gradient, the oscillatory behavior of ECE represents the scatter-replenish cycle of the resonant electrons the kinetic forces and wave-particle interactions, and the unstable-stable cycle of the whistler waves. This nonlinear behavior of the whistler waves and RE distribution will be studied in future.

\section{Summary}
\label{sec:conclusions}

ECE is a powerful tool to diagnose the runaway electron dynamics in tokamaks, but the interpretation of the signal is much more complicated than the black body case, and a synthetic diagnostic model is required. In this paper an ECE synthetic diagnostic model for arbitrary electron distributions is presented, and the conditions of runaway electrons producing significant non-thermal ECE radiation are discussed. According to the shape of the RE tail distribution, it was found that runaway electrons can give remarkable emissions of ECWs, but little contribution to the absorption. In addition, the emission power depends on both the energy and the pitch angle distribution of RE. Using the experimental parameters from DIII-D low-density flattop scenarios, we find that the RE tail can give strong non-thermal ECE radiation in both X-mode and O-mode, and make the ECE spectrum flatter. These results by the first time give solid explanations of these observations in RE experiments, and have good agreements with DIII-D diagnostic results. Further analysis of the weight function of ECE suggests that most of the high frequency ECE comes from REs in the low-energy regime with large pitch angles, a unique feature from other diagnostics. In addition to runaway electron studies, the synthetic diagnostic model can also be used to study ECE signals from other scenarios such as cases with strong current drive through wave particle interaction.

The kinetic instabilities associated with the non-monotonic and anisotropic RE beam play an essential role in producing the non-thermal ECE radiation.
In our simulations of the QRE experiments, we find that the ECE power is weak unless kinetic instabilities are excited and scatter the low-energy runaway electrons. In addition, the ECE from the unscatterd RE tail, with scattering from kinetic instabilities turned off, is not much different from a Maxwellian distribution, and only the scattered RE tails show the interesting properties of non-thermal ECE described above. Therefore, the significant non-thermal ECE radiation, which is ubiquitous in RE experiments, can be used as an indication of RE kinetic instabilities, which can be applied to other tokamak devices and ITER. The scattering effects from kinetic instabilities can also be important in the study of runaway electrons generated in disruptions, especially in ITER.

The model can be further improved in several ways. In the calculation of ECW emissions and absorptions, we use a uniform profile for the RE distribution near the core. This part can be improved by extending the kinetic simulation model from 2D to 3D, and applying a bounce-average kinetic simulation code like CQL3D or LUKE. Furthermore, in the calculation of ECE power using the reciprocal method, we only trace a single ray along the major radius direction, and ignore  the diffraction effects. This shortcoming can be improved by using an ECEI model incorporated in a recently-developed Synthetic Diagnostic Platform (SDP)\cite{shi_synthetic_2016-1}, which calculates ray tracing within a 2D wavefront plane including the refraction and diffraction of the wave.

\ack

Chang Liu wants to thank Robert Harvey, Ernest Valeo, John Krommes, Donald Spong, William Heidbrink and Yong Liu for fruitful discussions. This work has received funding from
the Department of Energy 
under Grant No. DE-SC0016268 and DE-AC02-09CH11466.

\bibliographystyle{iopart-num}
\bibliography{runaway-ece}
		
\end{document}